\documentclass[runningheads,a4paper]{llncs}

\usepackage{amssymb}
\usepackage{graphicx}

\usepackage{url}
\urldef{\mailsa}\path|{alfred.hofmann, ursula.barth, ingrid.haas, frank.holzwarth,|
\urldef{\mailsb}\path|anna.kramer, leonie.kunz, christine.reiss, nicole.sator,|
\urldef{\mailsc}\path|erika.siebert-cole, peter.strasser, lncs}@springer.com|
\newcommand{\keywords}[1]{\par\addvspace\baselineskip
\noindent\keywordname\enspace\ignorespaces#1}

\usepackage{todonotes}

\graphicspath{{./}{./fig/}}

\usepackage{algorithm}
\usepackage{algpseudocode}

\usepackage{latexsym}
\usepackage{amssymb}
\usepackage{tabularx}
\usepackage{amsfonts}
\usepackage{listings} 

\newcommand{\imp}{\Rightarrow} 

\newcommand\nonofoot[1]{%
   \begingroup
   \renewcommand\thefootnote{}\footnote{\kern-0.2ex#1}%
   \addtocounter{footnote}{-1}%
   \endgroup
}

\begin{document}

\mainmatter  

\title{Towards a better understanding and behavior recognition of inhabitants in smart cities.\newline
       A~public transport case}

\titlerunning{Towards a better understanding and behavior recognition of inhabitants...}

%
%
 \author{Rados{\l}aw Klimek \and Leszek Kotulski}
 \institute{AGH University of Science and Technology\\
 Al. Mickiewicza 30, 30-059 Krak\'ow, Poland\\
 \email{\{rklimek,kotulski\}@agh.edu.pl}
}
\authorrunning{R. Klimek L. Kotulski}


%
%

\maketitle

\begin{abstract}
The idea of modern urban systems and smart cities requires monitoring and
careful analysis of different signals.
Such signals can originate from different sources and one of the most promising is the BTS,
i.e.\ base transceiver station, an element of  mobile carrier networks.
This paper presents the fundamental problems of elicitation, classification and
understanding of such signals so as to develop context-aware and pro-active systems in urban areas.
These systems are characterized by the omnipresence of computing which is
strongly focused on providing on-line  support to users/inhabitants of smart cities.
A method of analyzing selected elements of  mobile phone datasets through
understanding inhabitants' behavioral fingerprints to obtain smart scenarios for public transport is proposed.
Some scenarios are outlined.
A multi-agent system is proposed.
A formalism based on graphs that allows reasoning about inhabitant behaviors
is also proposed.

\end{abstract}

\section{Introduction}
\label{sec:introduction}

\nonofoot{This is a draft/accepted version of the paper:
R.~Klimek, L.~Kotulski: Towards a better understanding and behavior recognition of inhabitants in smart cities.
A~public transport case.
\emph{Proceedings of 14th International Conference on Arificial Inteligence
	and Soft Computing (ICAISC 2015), 14--18 June, 2015, Zakopane, Poland.}.
Rutkowski, Leszek and Korytkowski, Marcin and Scherer, 
Rafal and Tadeusiewicz, Ryszard and Zadeh, Lotfi A. and Zurada, Jacek M. (Eds.),
ser.\ Lecture Notes in Artificial Intelligence,
vol.\ 9120,
pp.\ 237--246.
Springer Verlag.
\textcopyright~Springer 2015.
Available at: \texttt{DOI:10.1007/978-3-319-19369-4\_22} or \texttt{http://link.springer.com/chapter/10.1007\%2F978-3-319-19369-4\_22}}

We face today an unprecedented interest for the idea of smart cities.
This idea requires  smart analysis of many signals and information bits
which are generated in urban areas,
as well as the use of  network facilities and
 interaction of citizens through new technologies.
Thus,
new and innovative ways of analyzing behaviors in cities through understanding
the data they generate are needed.
New ways to analyze and classify this data,
as well as further reasoning,
in order to better understand and plan pro-active support offered by systems to inhabitants
are a crucial necessity.
Pervasive computing is an idea which assumes the omnipresence of computer systems
to give strong support for inhabitants in smart cities.
These systems must be characterized by  context-awareness,
basing on different urban signals, to provide pro-active assistance for
inhabitants.

Widespread availability and use of mobile phones,
as well as their growing ubiquity,
is based on  BTS wireless networks
which guarantee basic communication in the system.
Wireless networks have great potential to provide information
to identify activities of people.
People keep a phone with them most of the time.
Inhabitant movements and locations are being recorded in many different ways.
BTSs are responsible for communicating with mobile phones within the network.
They record many important and useful events,
stored in the CDR format,
for example about the presence of a phone device, which gives an
indication of the geographic location of the user.
Thus,
it allows to identify places of inhabitants' lives using mobile networks.
Being able to identify users' movements is crucial for smart decisions
as well as to work fast and get results in a short time.

We show that analyzing selected information generated by BTS devices
can indeed identify inhabitants' behaviors, help understand human mobility and social patterns
and  implement smart scenarios for software systems.
A classification of sensed behaviors for applications that operate in a smart city is proposed.
Outlines of smart scenarios are provided.
A multi agent system is proposed.
A formalism which allows reasoning about inhabitant behaviors in the BTS network is proposed.

The topic of sensing and monitoring urban activities basing on mobile phone datasets seems
hot and relatively new.
In work by Calabresse et al.~\cite{Calabrese-etal-2011}
a real time monitoring system is described.
Buses and taxis,
as well as pedestrians movements,
are positioned providing urban mobility.
In work by Gonzalez et al.~\cite{Gonzalez-etal-2008}
trajectories of anonymized mobile phone owners are discussed.
Human trajectories are characterized by a high degree of both
temporal and spatial regularity.
Work by Isaacman et al.~\cite{Isaacman-etal-2011} proposes
clustering and regression-oriented techniques
supporting identification semantically-meaningful locations (home, work).
Work by Reades et al.~\cite{Reades-etal-2007} offers
a new way of looking at the city as a holistic and dynamic system.
Some experiments in explorations in urban data collection are discussed.

\section{Technical preliminaries}
\label{sec:preliminaries}

Systems for mobile communications (e.g.\ GSM or UMTS) are now well established.
There are many works introducing in the world of data communication procedures,
e.g.\ work~\cite{Horak-2007}.
Selected technical aspects of such system are briefly outlined below.

\begin{figure}[htb]
\centering
\includegraphics[width=7cm]{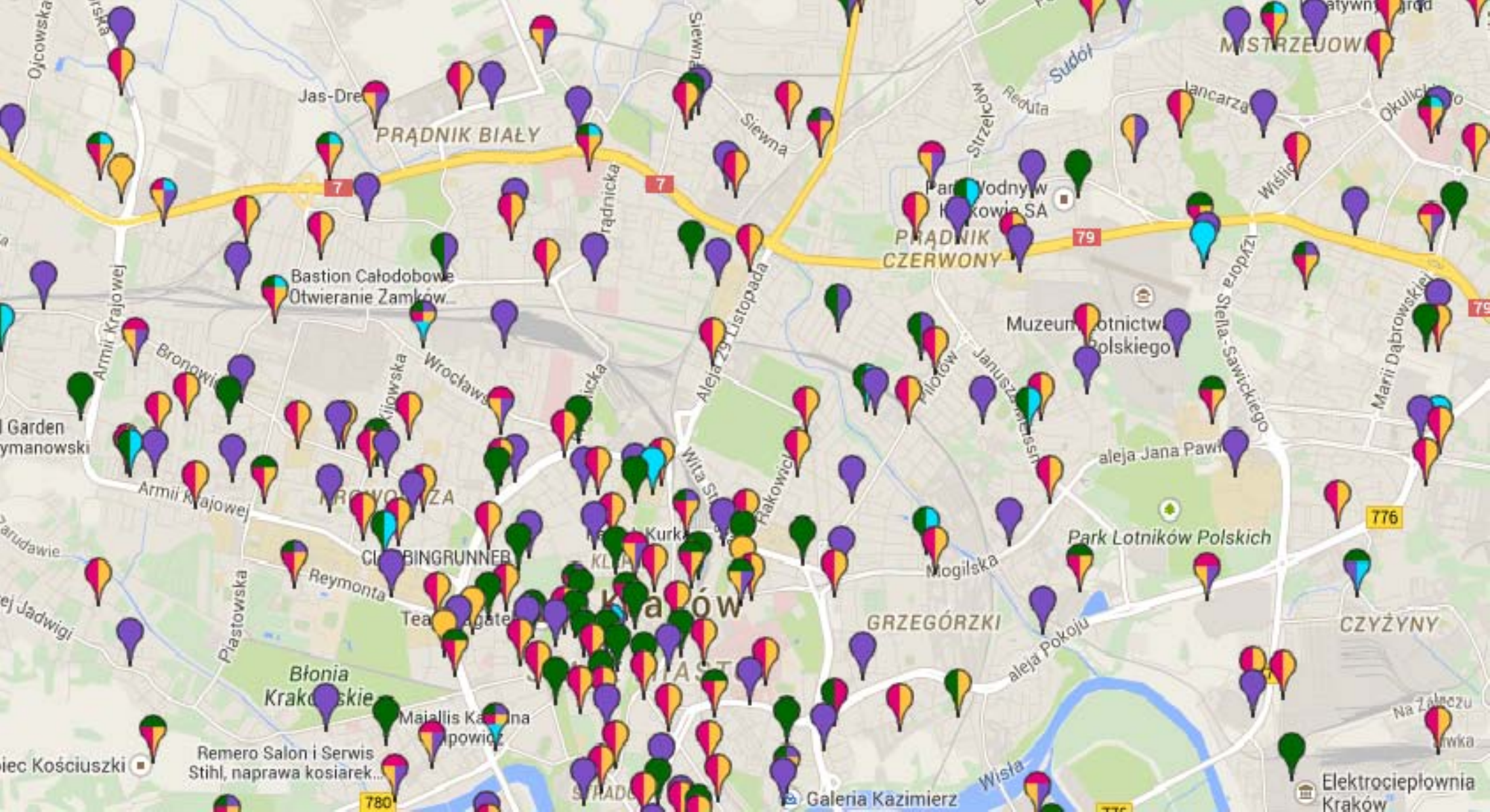}
\caption{A sample BTS city network (source: \texttt{http://btsearch.pl/})}
\label{fig:bts}
\end{figure}

The most obvious part of the cellular/mobile phone network is a base station.
A \emph{base transceiver station} (BTS)
is a piece of equipment that enables wireless communication between the user and the network.
Every BTS performs immediate communication with mobile phones.
Nowadays, cities and regions are covered with a relatively dense network of BTSs,
see for example Figure~\ref{fig:bts}.
Although outside the cities networks are less dense, in each case they gather and store important and interesting information about
users' activities.
Broadly speaking,
the entire network system consists of many elements that operate together but
an ordinary user is not aware of the different entities within the system.

A \emph{call detail record} (CDR) is contains data recorded and produced by
telecommunications equipment.
The purpose is to store information about current system usage;
however,
CDR is rather retrospective.
It contains data that is specific to a single instance of a phone call or other communication transaction.
The structure of CDR is relatively complex,
and its format varies among providers.
In some situations CDR can also be configured by a user.
There is an entry done for each call, at the start of a call and at the end of the call.
The management system is usually configured to update the CDRs periodically.
The CDR contains variables,
e.g.\ the called number. 
Variables might be grouped into:
variables used for identifying calls,
timestamps,
information related to signaling,
information related to media,
statistics,
information related to routing,
and others.
Records are very detailed and contain much information,
e.g.\ point of origin (sources),
points of destination (endpoints),
the phone number of the calling party,
the phone number of the  party being called,
duration of each call,
the amount billed for each call,
the route for a call entering the exchange,
the route for a call leaving the exchange,
call type (voice, SMS, etc.),
etc.
Some data depends on the service provider and even in a case of timestamps
there are over a dozen of different fields.

CDRs, as collections of information, have a special format~\cite{UK-standard-2014}.
Below is a sample fragment of a CDR text decoded from the binary format.
The first row must contain a header row which includes the field names:
\begin{verbatim}
"Call Type","Call Cause","Customer Identifier","Telephone Number Dialled",
"Call Date","Call Time","Duration","Bytes Transmitted","Bytes Received",
"Description","Chargecode","Time Band","Salesprice",
"Salesprice (pre-bundle)", "Extension","DDI","Grouping ID","Call Class",
"Carrier","Recording","VAT","Country of Origin","Network",
"Retail tariff code","Remote Network","APN","Diverted Number",
"Ring time","RecordID","Currency"
\end{verbatim}
The meaning of the columns is not analyzed here since they are
intuitive and the detailed discussion is outside the scope of the paper.
Below is a sample decoded text:
\begin{verbatim}
"V","0","+441999887000","+441999878333","28/01/2012","10:37:23",
"233","","","Hampton","UK Local","Peak","0.8","0.8","654",
"+441999887654","","UKL","Talk Talk","","S","","","","","","",
"","778789","GBP"

"VOIP","0","Brianb@M1.com","+442086019080","28/01/2012","10:39:23",
"345","","","On-Net","On-Net","Peak","0.0","0.0","","","","ON" ,
"Talk Talk","1","S","","","","","","","","8011229","GBP"
\end{verbatim}
There are many events that generate a CDR record,
e.g.\ data services, such as SMS and Internet access.
The gathered information allows to obtain BTS locations according to a mobile phone activity,
i.e.\ changing a location from one BTS to another.
Location information is extracted as part of the interaction data.
These location observations, i.e.\
\begin{itemize}
\item the moment of the phone's/object's entry into the area of a station (log in), and
\item the moment of they leave that area (log out),
\end{itemize}
are of fundamental importance to the
considerations given in the following sections of the paper.

\section{Behaviour recognition and classification}
\label{sec:classification}

This work discusses the possibility of analyzing
data generated and obtained from  BTS devices/stations.
Such stations constitute a rich source of information for smart and context-aware systems.
This information is related to many aspects of users'/inhabitants' behaviors
and base on relatively raw data.
From all the information generated by a BTS, the most important for these
considerations are events describing the presence/location (login/logout) of the phone in the BTS area.

\begin{figure}[htb]
\centering
\includegraphics[width=6cm]{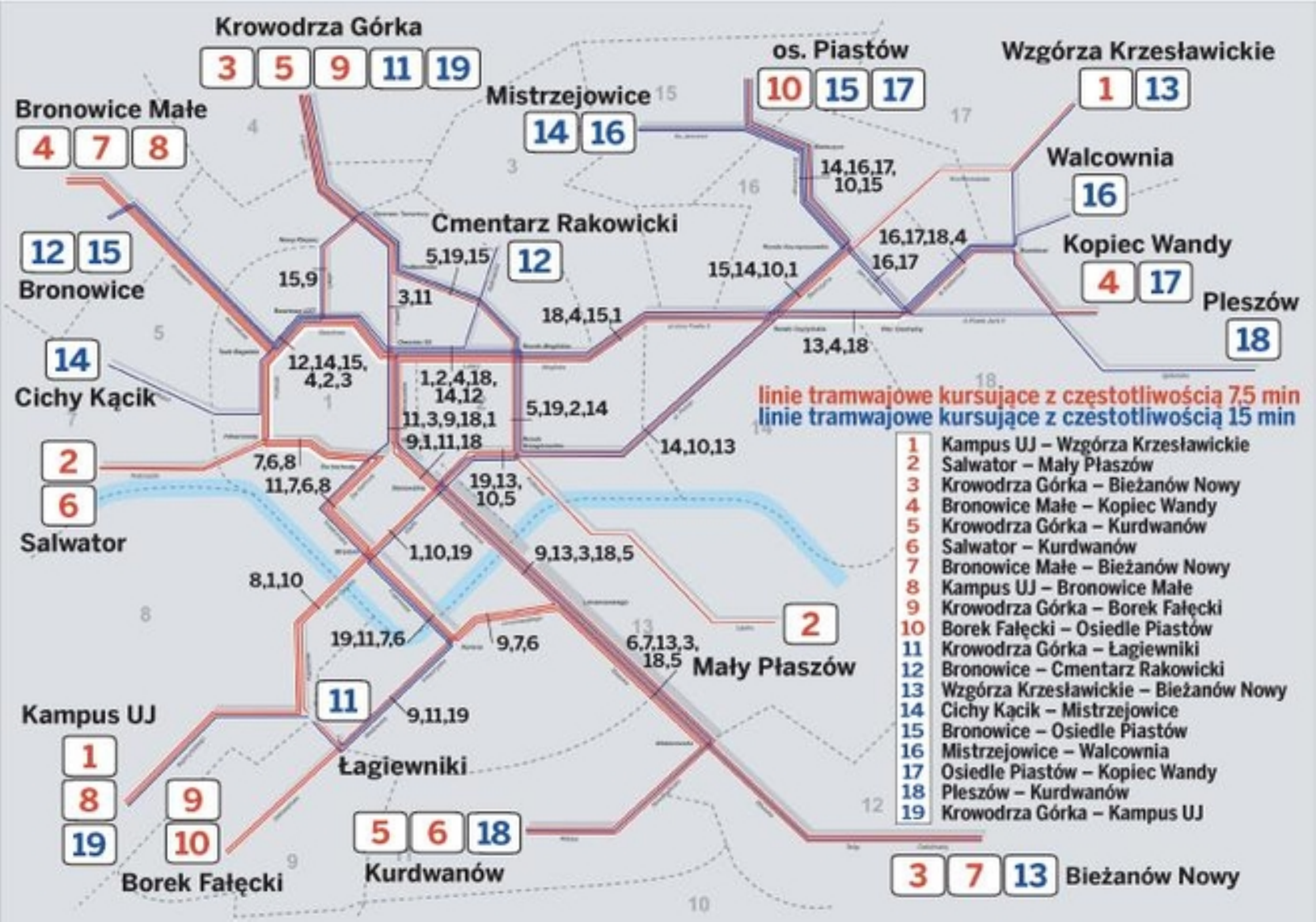}
\includegraphics[width=6cm]{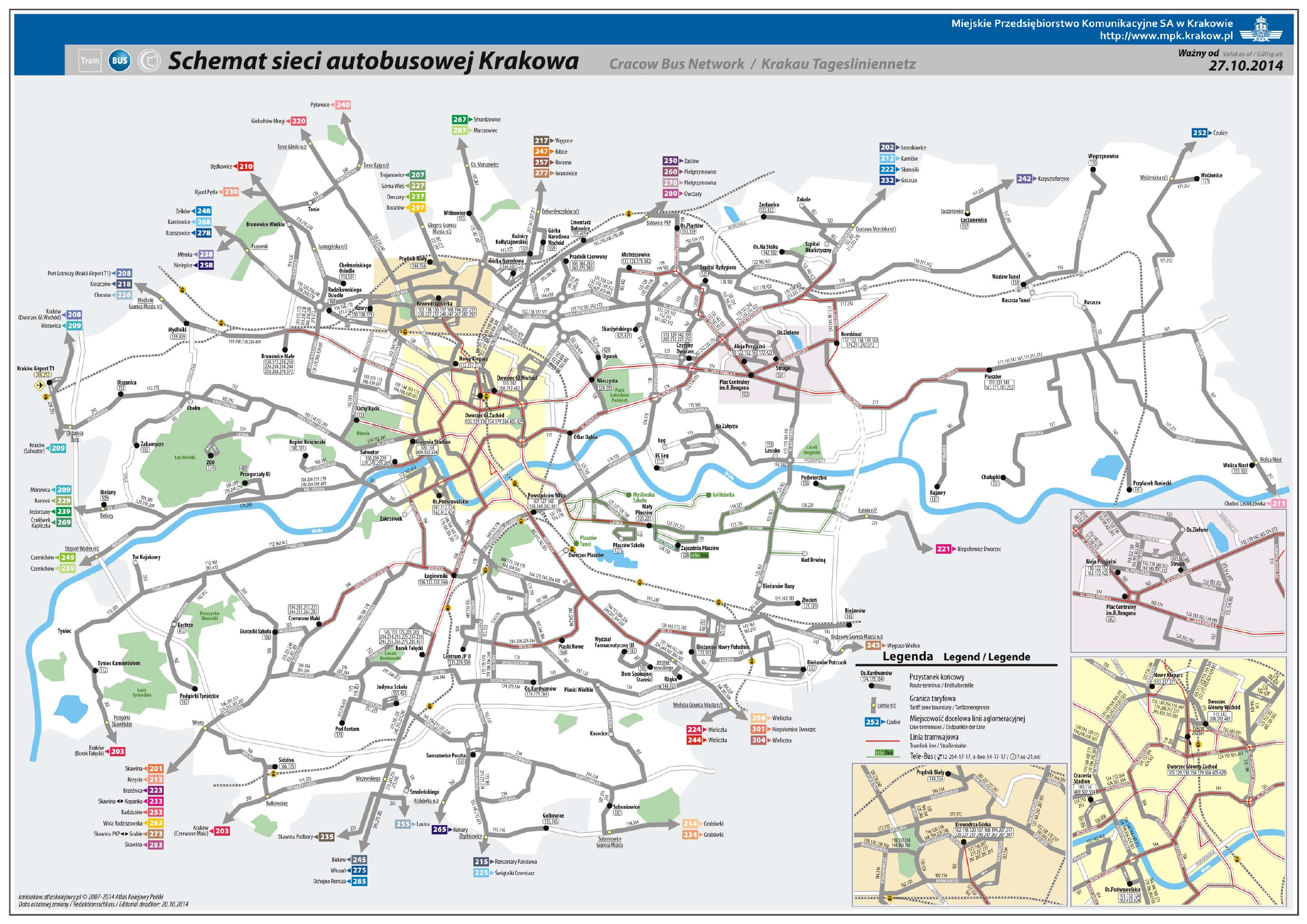}
\caption{A sample city tram/bus network (source: \texttt{http://www.mpk.krakow.pl/pl/
mapki-komunikacyjne/})}
\label{fig:tram-bus}
\end{figure}

Information, or events, obtained from the BTS network can be used to provide
the following classification of user behaviors:
\begin{enumerate}
\item \textbf{Static behavior},
       that is without moving outside the BTS area.
       Some scenarios which are appropriate for such behavior are proposed.
       The aim of such scenarios is to increase
       the user's/inhabitant's comfort of staying in a particular area.
       Appropriate algorithms can take  the behaviors registered in the past  into account
       which can build a suitable preference model~\cite{Klimek-Kotulski-2014-IE-AITAmI,Klimek-2013-icmmi}.
       For example,
       providing information about local customer services,
       shops or special offers.
       If preferences (behavior in the past) also also taken into account,
       then support for users/inhabitants staying in the area becomes more mature and valuable.
       For example,
       people working in local offices,
       when approaching the habitual and observed time for their lunch,
       are notified about current opportunities in their neighborhood,
       and table bookings in restaurants are suggested,
       offering personalized ads,
       etc.
       In other words,
       these actions are performed after gaining
       a deep understanding of the inhabitants' needs.

       The algorithm to identify home/residential or work/office locations might be based on
       the tracking the entire mobile activity during the selected days of a week and selected time of a day.
       In other words,
       home is defined when the mobile phone location is registered after
       a certain time at night during certain days.

       Another example of such support might be the situation
       when two people are informed about the possibility of their meeting
       as a result of being in the same geographic location,
       if such a meeting has been ``ordered'' before
       (e.g.:
       when I am in the same area as the person X,
       please notify and make an appointment).
\item \textbf{Dynamic, or mobile, behavior},
       that is related to the movement of both individual users or groups of inhabitants,
       between neighboring BTS points.
       It seems that dynamic behaviors, as understood here,
       give a great number of possibilities to introduce pro-active scenarios.
       Such scenarios are especially important for the ideas of pervasive computing
       and smart cities.
       The desired effect is particularly evident when
       applying some additional, and free, technologies
       related to the geographical location and maps,
       e.g.\ \emph{OpenStreetMap} OSM~\footnote{OSM is a project to create a free editable map of the world.},
       or maps of existing urban infrastructure networks,
       e.g.\ public transport lines, c.f.\ Fig.~\ref{fig:tram-bus}.
       Dynamic behaviors, due to their great potential for interesting uses, are to be discussed separately.
\end{enumerate}



Scenarios related to  dynamic behaviors are of fundamental importance in the paper.
Further considerations are focused on the classification of different types of
travel in the urban area.
The purpose of this classification is to distinguish two kinds of situations
which relate to the observed quick move:
\begin{enumerate}
\item a group of people traveling by public transport,
      i.e.\ simultaneously traveling groups of persons (phones) after finding
      that this is not a solitary case of traveling by private cars --
      confirmation of this case is a result of the following observations:
      the comparison to a similar behavior in the past,
      quick change of BTS areas, i.e.\ switching between BTSs,
      and  comparison of the current travel route
      with public transport lines, c.f.\ Fig.~\ref{fig:tram-bus};
\item people traveling by private cars --
      evidence of this case might be a result of the following observations:
      a greater speed of a travel comparing groups traveling by public transport,
      traveling outside the area of public transport lines, etc.
\end{enumerate}

When the above basic classification is done,
the following support for inhabitants  is considered as a result of a performance of
the context-aware and pro-active system:
\begin{enumerate}
\item (group) trip by public transport
\begin{itemize}
      \item finding convenient transfers for travelers if transfers are expected;
      \item in the case of transfers with a long wait for a new connection:
            finding  bar/cafeteria facilities in the area in advance to make a reservation;
      \item propose to notify people at home (destination) about the arrival time,
            or notifying of the planned arrival in advance a certain number of minutes before;
      \item finding alternative connections and transfers,
            if there are traffic jams which slow down a trip or make it difficult;
      \item notification of friends/colleagues about a common trip in the same vehicle of public transport,
            this fact can be confirmed by on-line analysis of social networks (e.g.: Facebook, Instagram, etc.);
      \item some others;
      \end{itemize}
\item (individual) trip by a car
      \begin{itemize}
      \item propose to notify people at home about the arrival time,
            or notifying of the planned arrival in advance a certain number of minutes before;
      \item warning regarding the approaching a critical locations,
            schools, places, crossroads, etc.,
            this service requires gathering additional data form
            OpenStreetMap;
      \item some others.
      \end{itemize}
\end{enumerate}

\section{System architecture}
\label{sec:system-agentowy}

In this section we introduce the agent system structure that
supports the IoT services mentioned in the previous sections.
Let us consider the structure of a system supporting
the simple task of determination of the way in which the owner of a mobile phone travels,
that is whether travel is done via public transportation or via private car.

\begin{figure}[htb]
\centering
\includegraphics[width=9cm]{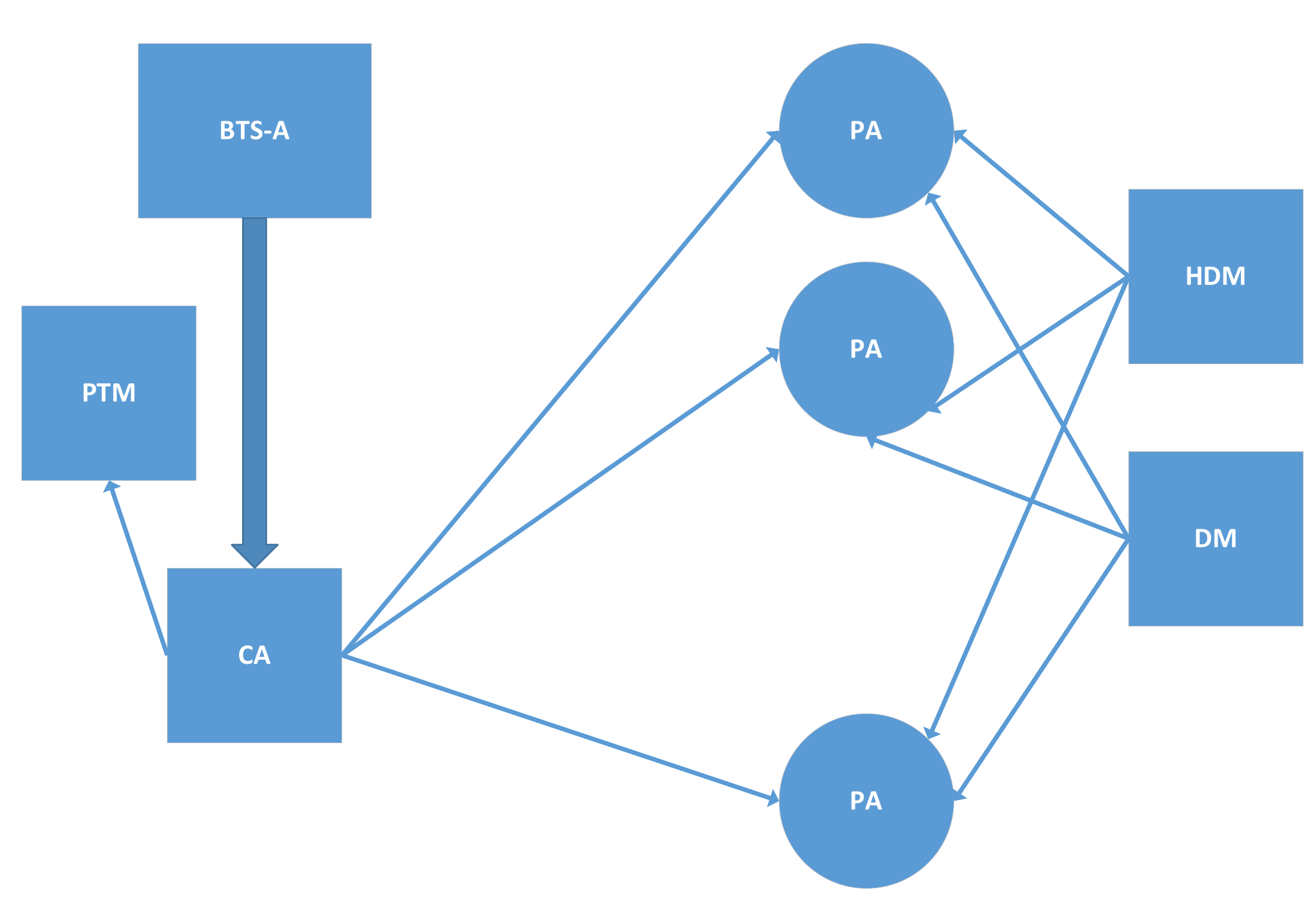}
\caption{A sample agent system for a public transportation system}
\label{fig:agent-system}
\end{figure}

The outline of the proposed agent system is shown in Fig.~\ref{fig:agent-system}.
The basic types of agents are:
\begin{itemize}
\item a \emph{personal agent} PA,
      that maintains the trace of the current route of a mobile phone in the entire city among BTS areas,
      as well as its stored characteristic, taking into account
      some historical information about previous behaviors;
      every mobile phone owner has its personal agent PA (informally, a guardian angel~PA);
\item  a \emph{BTS agent} BTS-A for every base transceiver station BTS,
       that, for every personal agent PA that came into the BTS area
       at the same time slot,
       creates a temporary \emph{coordination agent} CA which
       stamps the time of the entry to an area and
       try to characterize every PA's behaviour in the following way:
\begin{enumerate}
\item if jump among (at least three) regions is slow, then walking,
\item if jump among (at least three) regions is medium, then private car or public transportation,
\item if jump among (at least three) regions is fast, then private car;
\end{enumerate}
the speed (slow, medium, fast) is determined taking into account speeds observed in the considered area;
every agent CA is removed/killed when its reasoning process,
initiated by an agent BTS-A for a list of jumping PA agents passed to CA is finished;
\item a \emph{public transportation manager agent} PTM\footnote{that can be associated with some buses,
                if such buses are identified in the network,
                or of it represents a virtual bus -–
                a group of personal agents that move together in the same destination}
      that tries to recognize and represent the group of personal agents PAs that move in the same line.
\end{itemize}

The following are rules for the CA agent creation algorithm:
\begin{itemize}
\item for every pair of neighboring BTS regions
      the BTS generates a list of PA agents which passed/jumped
      between two regions in a given time slot;
\item the CA agent is created and the list of jumping agents PAs constitute its input;
      CA gathers information about the trace of previous travel and creates
      a \emph{travel graph} for all PA agents considered by CA.
\end{itemize}

Let us consider a graph $G=\langle V,E \rangle$,
c.f.\ \cite{Kotulski-2008-iccs},
where vertices $V$ are parts of the BTS state which
maintains the collection of PA agents that jump to
this BTS from other BTS at a certain time,
and $E$ shows the movement between nodes,
i.e.\ $(v,w) \in E$.
Nodes in the travel tree represents the BTS in time $t$,
so we will describe it as a pair $(BTS_{ID}, t)$ --
such node  will be called node at  level $t$.
Initially, at time $t$,
there is only one node for which the CA agent has been created.
At level $t-1$, there are nodes from which an object jumps to nodes at level $t$.
Edges show from which node at level $t-1$ an object is moved to a node at level $t$.
Each edge is labelled by the name of the moving object.

Let us notice that the travel graph is a multi-edge graph,
which means that more than one edge  can exist between two nodes.
Next we will designate the object using a recursive algorithm;
at  level $t-1$  we initially assume that with node $x = (id,t-1)$
there are associated all objects that will jump from $x$ to any node at the higher level.
For each object $Q$ associated with node $x$, agent CA retrieves the information about
the previous traversal of $Q$  from agent $PA_Q$;
It should be noticed that:
\begin{itemize}
\item the number of associated object can grow,
      because agent $PA_Q$  can remember that at time  $t-1$, $Q$ was in $id_{BTS}$ with other objects;
\item this travel enriches the travel graph at levels lower then $t-1$.
\end{itemize}
When we gather all the information about the route of the object from
level $t-1$,  we will update the information about the nodes at  level $t-2$ and the following ones.
Time is an attribute of the edge.
The following rules should be fulfilled in the travel graph:
\begin{itemize}
\item $\forall v,w \in V: (w,PA_Q) \in E \imp time(v)+1 = time(w)$;
\item $y_v=x_w$;
\item and agent $PA_Q$ moves from $w$ to $v$.
\end{itemize}

The travel graph is a multi-graph, which means that  more than one edge can exist between the same two graphs nodes.
They are differentiated by a label that identifies the PA agent.
This graph might constitute a base for reasoning.

The decision made by the CA agent is supported by the information from
the PTM agent that can verify if the route traveled by a PA agent can be covered using public transportation.
Let us note that we  still have a problem with differentiation of two situations:
\begin{enumerate}
\item traveling by a bus,
\item traveling in a column of a few/column cars.
\end{enumerate}
In such case,  historical data is used to make a decision with
the most probability. Let us note that in  next steps we can determine
the type of transportation because either it is not possible
to find a bus travel in this destination or the column of cars has been split.
\begin{figure}[htb]
\centering
\includegraphics[width=10cm]{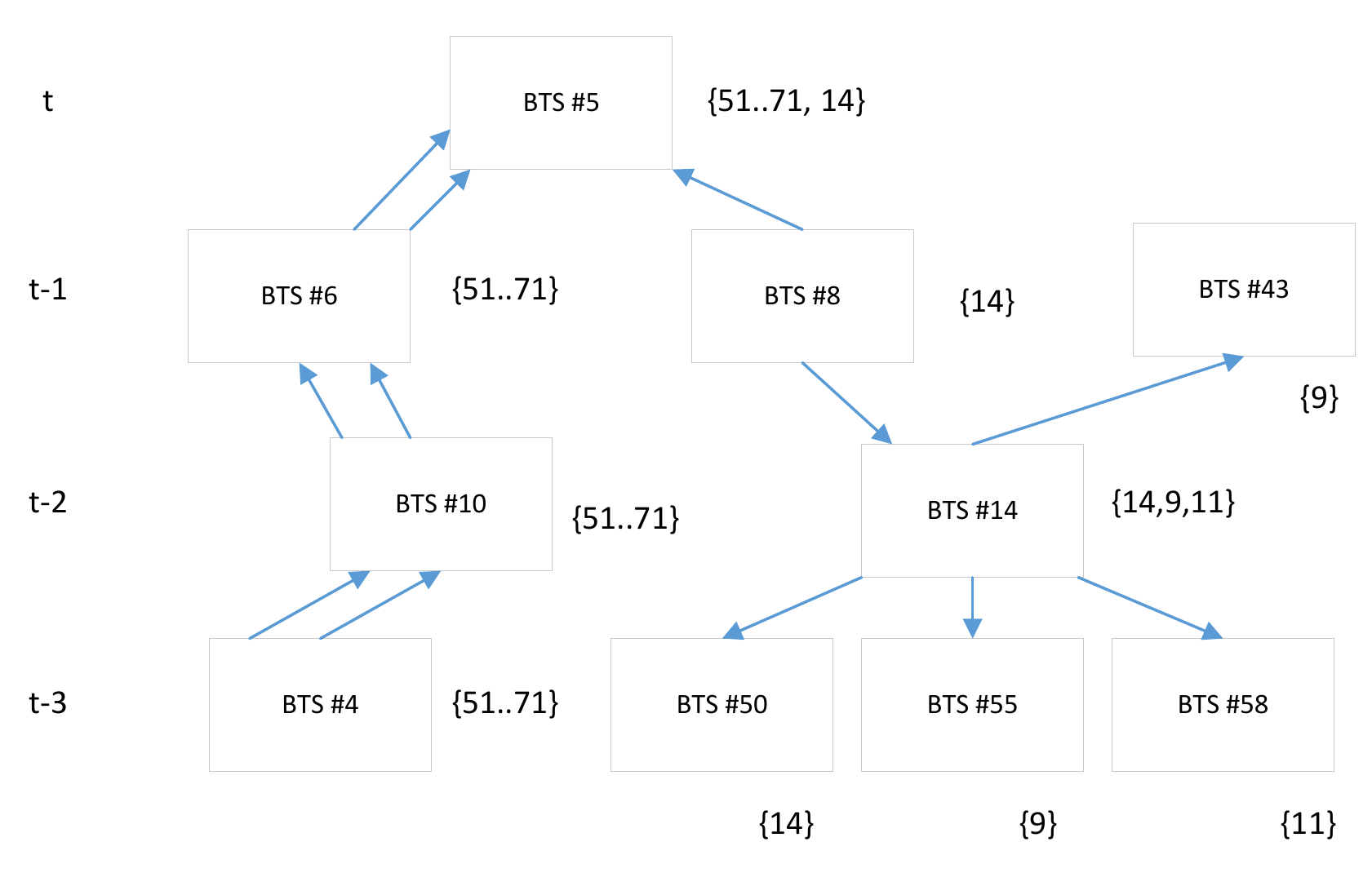}
\caption{A screen shot of a sample travel graph}
\label{fig:travel-graph}
\end{figure}
A~sample travel graph is shown in Figure~\ref{fig:travel-graph},
where two arrows from one node to another
(if any)
symbolize that more than one edge  exists between two nodes,
(i.e.\ there are two or more edges).
Analysis of the graph shows that
agents from 51 to 71 are traveling together using the public transport system.
Agent 14 travels by car.

The agent system can be extended by two more types of agents:
\begin{itemize}
\item a \emph{historical data maintainer agent} HDM, i.e.\
       an agent that maintains the historical data of personal agents;
\item a \emph{data mining agent} DM,
      that takes the trace of route traveled by a personal agent and processes it
      into some interesting historical behavior; for example:
\begin{itemize}
\item routes that are covered by walking, via public or private transportation,
\item public agents which usually travel together,
\item a schedule of routes that are executed periodically.
\end{itemize}
\end{itemize}
A personal agent PA returns the information to
the DM agent (after finishing the travel) and
takes it from HDM before starting a new travel.

Historical data maintained by some agents
open an interesting issue that supplements the approach presented here.
The historical behaviors are encoded into logical specifications,
and can be later analyzed for satisfiability,
c.f.\ works~\cite{Klimek-Kotulski-2014-IE-AITAmI} 
or~\cite{Klimek-2014-AMCS,Klimek-Faber-Kisiel-Dorohinicki-2013-fedcsis,Klimek-Rogus-2014-ICAISC},
supporting the current reasoning process and behavior recognition.


\section{Conclusions}
\label{sec:conclusions}

In this paper, the problem of sensing inhabitant behaviors in a smart city are considered.
The classification of behaviors observed using the BTS networks is proposed.
A public transportation case is discussed, and a multi-agent system is proposed.
This work opens a research area which is of crucial importance for
the idea of smart cities.

Future works may include the implementation of the reasoning engine.
It should result in a CASE software,
which could be a first step involved in creating industrial-proof tools.

\bibliographystyle{splncs03}
\bibliography{bib-rk,bib-rk-main,bib-rk-pervasive,bib-rk-graph}

\begin{thebibliography}{10}
\providecommand{\url}[1]{\texttt{#1}}
\providecommand{\urlprefix}{URL }

\bibitem{Calabrese-etal-2011}
Calabrese, F., Colonna, M., Lovisolo, P., Parata, D., Ratti, C.: Real-time
  urban monitoring using cell phones: A case study in rome. IEEE Transactions
  on Intelligent Transportation Systems  12(1),  141--151 (2011)

\bibitem{UK-standard-2014}
{Federation of Communication Services}: {UK Standard for CDRs. Standard CDR
  Format} (January 2014)

\bibitem{Gonzalez-etal-2008}
Gonzalez, M.C., Hidalgo, C.A., Barabasi, A.L.: Understanding individual human
  mobility patterns. Nature  453(7196),  779--782 (June 2008)

\bibitem{Horak-2007}
Horak, R.: Telecommunications and Data Communications Handbook.
  Wiley-Interscience (2007)

\bibitem{Isaacman-etal-2011}
Isaacman, S., Becker, R., Cáceres, R., Kobourov, S., Martonosi, M., Rowland,
  J., Varshavsky, A.: Identifying important places in people’s lives from
  cellular network data. In: Lyons, K., Hightower, J., Huang, E. (eds.)
  Pervasive Computing, Lecture Notes in Computer Science, vol. 6696, pp.
  133--151. Springer Berlin / Heidelberg (2011)

\bibitem{Klimek-2013-icmmi}
Klimek, R.: Preference models and their elicitation and analysis for
  context-aware applications. In: Gruca, A., Czach\'{o}rski, T., Kozielski, S.
  (eds.) Proceedings of 3rd International Conference on Man-Machine
  Interactions (ICMMI 2013), 22--25 October 2013, Brenna, Poland. Advances in
  Intelligent Systems and Computing, vol. 242, pp. 353--360. Springer
  International (2014)

\bibitem{Klimek-2014-AMCS}
Klimek, R.: A system for deduction-based formal verification of
  workflow-oriented software models. International Journal of Applied
  Mathematics and Computer Science  24(4),  941--956 (2014),
  \url{http://www.amcs.uz.zgora.pl/?action=paper\&paper=802}

\bibitem{Klimek-Faber-Kisiel-Dorohinicki-2013-fedcsis}
Klimek, R., Faber, {\L}., Kisiel-Dorohinicki, M.: Verifying data integration
  agents with deduction-based models. In: Proceedings of Federated Conference
  on Computer Science and Information Systems (FedCSIS 2013), 8--11 September
  2013, Krak\'{o}w, Poland. pp. 1049--1055. IEEE Xplore Digital Library (2013)

\bibitem{Klimek-Kotulski-2014-IE-AITAmI}
Klimek, R., Kotulski, L.: Proposal of a multiagent-based smart environment for
  the iot. In: Augusto, J.C., Zhang, T. (eds.) Workshop Proceedings of the 10th
  International Conference on Intelligent Environments, Shanghai, China, 30th
  June--1st of July 2014. Ambient Intelligence and Smart Environments, vol.~18,
  pp. 37--44. IOS Press (2014)

\bibitem{Klimek-Rogus-2014-ICAISC}
Klimek, R., Rogus, G.: Modeling context-aware and agent-ready systems for the
  outdoor smart lighting. In: Rutkowski, L., Korytkowski, M., Scherer, R.,
  Tadeusiewicz, R., Zadeh, L.A., Zurada, J.M. (eds.) Proceedings of 13th
  International Conference on Arificial Inteligence and Soft Computing (ICAISC
  2014), 1--5 June, 2014, Zakopane, Poland. Lecture Notes in Artificial
  Intelligence, vol. 8468, pp. 269--280. Springer Verlag (2014)

\bibitem{Kotulski-2008-iccs}
Kotulski, L.: Gradis -- multi-agent environment suppporting distributed graph
  transformations. In: Proceedings of 8th International Conference on
  Computational Science (ICCS 2008), 23--25 June 2008, Krakow, Poland. Lecture
  Notes in Computer Science, vol. 5103, pp. 644--653. Springer Verlag (2008)

\bibitem{Reades-etal-2007}
Reades, J., Calabrese, F., Sevtsuk, A., Ratti, C.: Cellular census:
  Explorations in urban data collection. IEEE Pervasive Computing  6(3),
  30--38 (2007)

\end{thebibliography}

\end{document}